\documentclass[twocolumn,showpacs,superscriptaddress,floatfix,preprintnumbers,amssymb,amsmath]{revtex4}

\usepackage{graphicx}
\usepackage{dcolumn}
\usepackage{bm}
\usepackage[latin1]{inputenc}
\usepackage[mathscr]{eucal}
\usepackage{epsfig}
\usepackage{rotating}

\begin{document}

\title{Suppression of the critical current of a balanced SQUID}

\author{Antti Kemppinen}
\affiliation{Centre for Metrology and Accreditation (MIKES), P.O. Box 9, 02151 ESPOO, Finland}
\author{Antti J. Manninen}
\affiliation{Centre for Metrology and Accreditation (MIKES), P.O. Box 9, 02151 ESPOO, Finland}
\author{Mikko M\"{o}tt\"{o}nen}
\affiliation{Low Temperature Laboratory, Helsinki University of
Technology, P.O. Box 3500, 02015 TKK, Finland}
\affiliation{Laboratory of Physics, Helsinki University of Technology,
P.O. Box 5100, 02015 TKK, Finland}
\author{Juha J. Vartiainen}
\affiliation{Low Temperature Laboratory, Helsinki University of
Technology, P.O. Box 3500, 02015 TKK, Finland}
\author{Joonas T. Peltonen}
\affiliation{Low Temperature Laboratory, Helsinki University of
Technology, P.O. Box 3500, 02015 TKK, Finland}
\author{Jukka P. Pekola}
\affiliation{Low Temperature Laboratory, Helsinki University of
Technology, P.O. Box 3500, 02015 TKK, Finland}

\begin{abstract}
We present an experimental study of the magnetic flux dependence of the critical current of a 
balanced SQUID with three Josephson junctions in parallel. Unlike for ordinary dc SQUIDs, the 
suppression of the critical current does not depend on the exact parameters of the Josephson
junctions. The suppression is essentially limited only by the inductances of the SQUID loops. We 
demonstrate a critical current suppression ratio of higher than 300 in a balanced SQUID with a 
maximum critical current 30~nA.
\end{abstract}


\maketitle Dc SQUIDs are routinely used to provide tunable critical current $I_\mathrm{c}$, e.g., in 
quantum computing applications~\cite{nakamura, makhlin}. For example, some 
charge qubits and charge pumps would benefit if $I_\mathrm{c}$ could be tuned very close to
zero~\cite{faoro,niskanenpumppu,cholacomb}. These devices require Coulomb blockade, and 
hence very small Josephson junctions (JJ) must be used. The range of the critical current of a dc 
SQUID is $|I_{\mathrm{c}1}\pm I_{\mathrm{c}2}|$ where $I_{\mathrm{c}i}$ are the critical currents 
of the individual JJs. It is not possible to fabricate two identical junctions, hence some 
residual critical current always exists. As suggested in
Refs.~\cite{faoro,niskanenpumppu,cholacomb}, this problem can be completely 
eliminated by using a balanced SQUID, i.e.~a structure with three parallel JJs in two superconducting 
loops with individual magnetic flux controls. Surprisingly only few experiments have utilized 
individual on-chip flux controls instead of a homogeneous external flux (see
e.g.~Refs.~\cite{Niskanen2005,himescience}). In this paper, we present to our knowledge the first 
experimental study of the balanced SQUID. Our prime motivation to search for high critical current 
suppression ratio is to improve the accuracy of the Cooper pair sluice~\cite{niskanenpumppu}. It is a 
current pump that can produce a current as high as 1~nA~\cite{Niskanen2005, aplpumppu}, but whose 
accuracy is not yet sufficient for a quantum current standard.

\begin{figure}[h]
    \begin{center}
    \includegraphics[width=.47\textwidth]{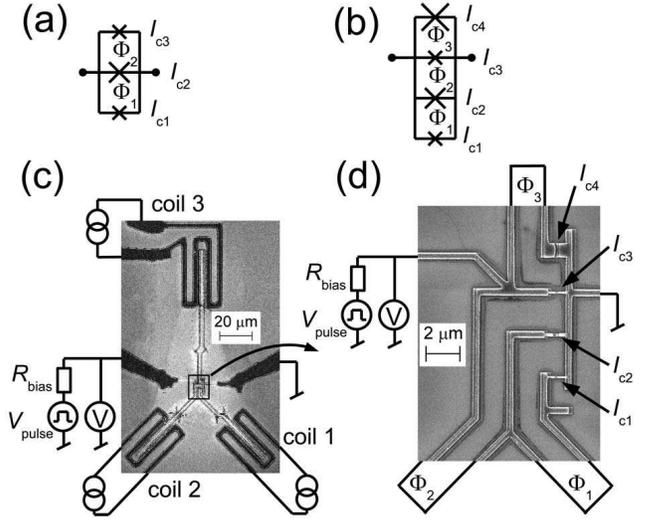}
    \end{center}
   \caption{\label{fig:sample} (a) Balanced SQUID. The middle junction is larger than the
   others.
   (b) Large detector junction in parallel with the balanced SQUID.
   (c) Scanning electron micrograph of the sample showing the on-chip coils and the narrow SQUID 
   loops, and a simplified sketch of the measurement setup. Here the resistance in series with the 
   pump is $R_\mathrm{bias}=100~\textrm{k}\Omega$.
   (d) Magnified view of the junctions and a sketch of the SQUID loops.}
\end{figure}

A schematic picture of the balanced SQUID is presented in Fig.~\ref{fig:sample}a. It or any other
system of $n$ JJs in parallel forming $n-1$ loops can be modeled as follows. The 
total current is the sum of the currents through the individual JJs, and the vector of the JJ 
currents is $\mathbf{I}=\mathbf{I}_\mathrm{c}\sin\bm{\phi}$, where $\mathbf{I}_\mathrm{c}$ is a 
diagonal matrix of the critical currents $I_{\mathrm{c}i}$ of the JJs. A vector containing the sines 
of the phases $\phi_i$ over the JJs is denoted by $\sin\bm{\phi}$. The effect of the magnetic flux
$\Phi_1$ on loop 1 is
\begin{equation}\label{1vuo}
\phi_1-\phi_2=\frac{2\pi}{\Phi_0}(\Phi_1
-L_{11}I_{\mathrm{c}1}\sin\phi_1+L_{12}I_{\mathrm{c}2}\sin\phi_2),
\end{equation}
where $\Phi_0$ is the flux quantum. The magnetic flux induced to the loop by the currents flowing in 
the branches $j$ of the structure are proportional to the inductances $L_{1j}$. For several loops, 
Eq.~(\ref{1vuo}) can be written in a matrix form: $\mathbf{M}\bm{\phi}=\mathbf{f}-\mathbf{LI}$. Here,
$\mathbf{f}=(2\pi/\Phi_0)\mathbf{\Phi}$, where $\mathbf{\Phi}$ is a vector of the fluxes of the loops.
The $(n-1)\times n$ inductance matrix $\mathbf{L}$ contains the factor $2\pi/\Phi_0$.

In the limit of zero inductances, the total current is obtained in an analytic form
$I=I_{\mathrm{c}1}\sin(\phi_1)+I_{\mathrm{c}2}\sin(\phi_1-f_1)+\ldots+I_{\mathrm{c}n}\sin(\phi_1-f_1-\ldots -f_{n-1})$, where $f_i=2\pi\Phi_i/\Phi_0$. The critical current of the whole structure
$I_\mathrm{c}(f_1,\ldots ,f_{n-1})$ as a function of the magnetic fluxes can be solved with the help 
of trigonometric identities. The full model including the 
inductances requires extensive computation over the phases of the JJs, whereas the analytic model 
allows direct calculation of the critical current. Therefore, the analytic zero-inductance model 
is helpful for the analysis and parameter fitting of the measurement results when the inductances 
are small.

Since we are interested in the suppression of a critical current that is originally of the order of 
tens of nanoamperes, we must be able to measure the critical current at least at 0.1~nA level. So low 
currents belong to the regime of phase diffusion, where a current-voltage ($IV$) measurement gives a
maximum supercurrent roughly proportional to $I_\mathrm{c}^2$. Moreover, the result is 
very sensitive to the electromagnetic environment~\cite{inggra}. Even an $IV$ measurement of a 
critical current of about 50~nA requires a special environment~\cite{steinbach}. Therefore we add a 
detector junction with $I_{\mathrm{c}4}>100$ nA in parallel with our balanced SQUID, see
Fig.~\ref{fig:sample}b. This superconducting shunt protects the balanced SQUID from the environment. 
The critical current of the four-junction system exhibits sinusoidal modulation around
$I_{\mathrm{c}4}$ as a function of the flux $\Phi_3$. The amplitude of the modulation equals the 
critical current of the balanced SQUID.

We measure switching into the normal state to determine the critical current at a certain magnetic 
flux set $\bm{\Phi}$ by feeding current pulses of constant length and variable height 
through the device. Switching produces a voltage pulse. We measure the switching 
probability at typically 5 points and determine the current, $I_{50}$, at which the
system has a 50~\% probability to switch to the normal state. At the measurement temperature 125~mK, 
with $I_{\mathrm{c}4}\approx 200~\mathrm{nA}$ and with about 1~ms pulse lengths,
$I_{50}=\alpha I_\mathrm{c}$, where the factor $\alpha$ is between 0.6 and 0.7~\cite{kivioja}. To 
distinguish between the real critical currents and the measured values obtained from the switching 
experiments, we use the notation $i_{\mathrm{c}i}=\alpha I_{\mathrm{c}i}$ for the measured critical 
currents. Here the index $i$ refers to an individual JJ or to the balanced SQUID.

\begin{figure}[h]
    \begin{center}
    \includegraphics[width=.47\textwidth]{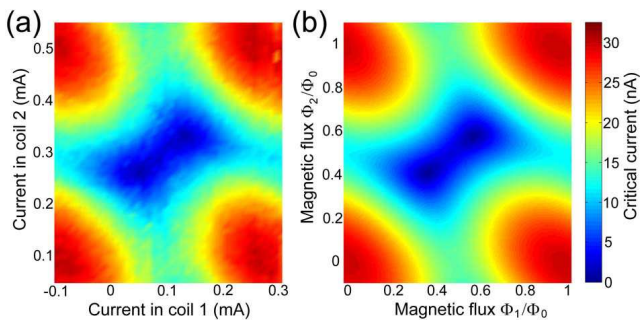}
    \end{center}
   \caption{\label{fig:vuokartat} (a) Measured critical current $i_\mathrm{cb}$ of the balanced SQUID 
   as a function of the coil currents. The maximum is shifted from zero current due to an offset flux.
   (b) Respective theoretical flux modulation of the critical current calculated with the 	
   parameters fitted from the measurement.}
\end{figure}

The measured flux modulation of the critical current is presented in Fig.~\ref{fig:vuokartat}a. For 
each data point of the 2D chart, we measured $I_{50}$ for 11 different currents in coil~3 mapping 
about one flux quantum in $\Phi_3$. The critical current $i_\mathrm{cb}$ of the balanced SQUID was 
extracted from the difference of the maximum and minimum critical current of the whole structure as a 
function of the current in coil~3.

The critical currents of junctions 1 and 3 are almost equal in our balanced SQUID design,
see Fig.~\ref{fig:sample}. Junction 2 is larger. The critical current of the SQUID is tunable 
to zero if $I_{\mathrm{c}2}\leq I_{\mathrm{c}1}+I_{\mathrm{c}3}$. The flux modulation of the 
SQUID is periodic in the square with $\Phi_1,\Phi_2\in [0,\Phi_0]$, see Fig.~\ref{fig:vuokartat}. 
The critical current has the maximum value at the corners of the square. In the middle region, there 
are two minima. If $I_{\mathrm{c}2}\geq I_{\mathrm{c}1}+I_{\mathrm{c}3}$, there would be only one 
minimum with value $I_{\mathrm{c}2}-(I_{\mathrm{c}1}+I_{\mathrm{c}3})$. The 
two minima are close to the line $(0,0)\to (\Phi_0,\Phi_0)$. In the Cooper pair sluice, one could use
a coil coupled symmetrically to the loops instead of the individual couplings exploited here. It 
would then be possible to move along the direction $(0,0)\to (\Phi_0,\Phi_0)$ from the minimum close 
to the maximum with a single rf control. An asymmetrically coupled dc controlled coil would also be 
required, but dc signals are easier to implement.

We fitted the zero-inductance model to the entire 3D flux modulation data. As fitting parameters, we 
used the critical currents of the junctions, $\mathbf{i}_\mathrm{c}$, the offset flux vector
$\mathbf{\Phi}_\mathrm{offset}$, and the matrix $\mathbf{K}$ depicting the couplings and cross 
couplings between the coils and the SQUID loops:
$\mathbf{\Phi}=\mathbf{\Phi}_\mathrm{offset}+\mathbf{KI}_\mathrm{coil}$. The fitted parameters were 
used to calculate the theoretical flux modulation chart presented in Fig.~\ref{fig:vuokartat}b.
The resulting critical currents are $i_{\mathrm{c}1}=7.4$~nA, $i_{\mathrm{c}2}=13.5$~nA,
$i_{\mathrm{c}3}=9.0$~nA, and $i_{\mathrm{c}4}=142.5$~nA. The maximum critical currents of the 
balanced SQUID and the whole structure are about 30~nA and 172~nA, respectively.

Next, we performed a measurement at $17\times 17$ points near the minimum of the critical current of 
the balanced SQUID with about $\Phi_0/400$ step in $\Phi_1$ and $\Phi_2$. Since the coupling of the 
detector coil 3 to the balanced SQUID loops is expected to have the most pronounced effect near the 
minimum, we fitted the terms $K_{13}$ and $K_{23}$ of the flux coupling matrix again to these data. 
The flux coupling matrix combined from the two fits is
\begin{equation}\label{Kmat}
\mathbf{K}=\left( 
  \begin{array}{ccc}
   2.52 & 0.09 & 0.10 \\
   0.04 & 2.39 & 0.11 \\
   0.05 & 0.06 & 2.63 \\
  \end{array}
\right) \; \Phi_0/\textrm{mA}.
\end{equation}
The cross coupling terms $K_{13}$ and $K_{23}$ are somewhat larger than expected, partly because 
the bonding wires of coil 3 pass near the sample. This enhances the effect that the fluxes $\Phi_1$ 
and $\Phi_2$ do not remain constant when the flux sweep over $\Phi_3$ is being performed.

\begin{figure}[h]
    \begin{center}
    \includegraphics[width=.47\textwidth]{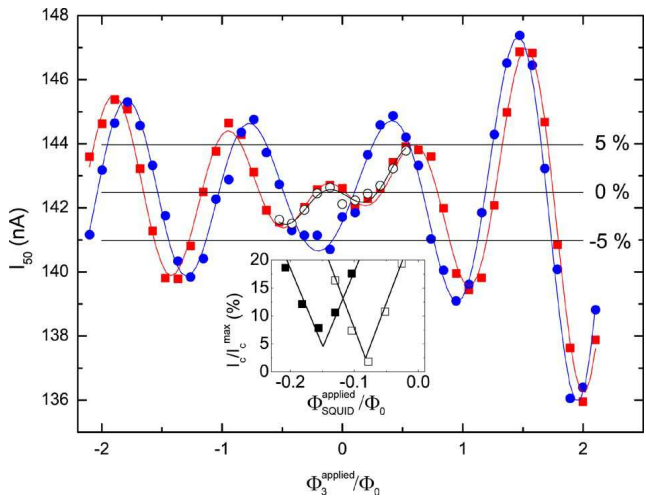}
    \end{center}
   \caption{\label{fig:minmodul} Detector flux modulations
   ($\Phi_3^\mathrm{applied}=K_{33}I_\mathrm{coil3}$) near the minimum of $i_\mathrm{cb}$. The 
   horizontal lines present change from $i_{\mathrm{c}4}$ as a percentage of the maximum of
   $i_\mathrm{cb}$. The open circles are an example 
   of the data used for the parameter fits of the cross coupling terms $K_{13}$ and $K_{23}$. The 
   black line is the corresponding curve of the zero-inductance model yielding a critical current
   $i_\mathrm{cb}=0.035~\mathrm{nA}$. The data spanning four flux quanta presented by the red squares 
   and the blue circles were not used for fitting. The corresponding red and blue lines are the 
   predicted flux modulations yielding $i_\mathrm{cb}=0.17~\mathrm{nA}$ and
   $i_\mathrm{cb}=1.8~\mathrm{nA}$, respectively. The inset shows typical results of similar critical 
   current measurements of two ordinary dc SQUIDs, expressed as a function of the applied magnetic 
   flux of the SQUID. The minima of the critical currents are 2.5~\% and 4.6~\% of the maxima, which 
   are about 30~nA for both SQUIDs.}
\end{figure}

One of the measured flux modulation curves with respect to the detector coil 3 is shown 
in Fig.~\ref{fig:minmodul} by open circles. The chosen curve corresponds to the measurement point
$(\Phi_1,\Phi_2)$ that is closest to the fitted minimum of the critical current of the balanced 
SQUID. Due to the cross coupling, the lowest critical currents cannot be directly extracted from the 
modulation amplitude. The zero-inductance model gives the modulation curve shown by the black line 
and a critical current 0.035~nA at this measurement point.

The parameter fits were performed to data extending over one flux quantum in the detector loop. To 
test the validity of our model, we extrapolated the fits over four periods in $\Phi_3$ and compared
to the measurements. The theoretical predictions and the measured data at two $(\Phi_1,\Phi_2)$ 
points near the minimum are shown in Fig.~\ref{fig:minmodul}. The only fitting parameter was the 
overall magnitude of each data set. The deviations of the measurement points from the theoretical 
curves are less than about 0.2~nA, and no systematics of the deviations is observed.

The remaining sources of residual critical current in the balanced SQUID are noise and the inductances 
of the loops. The effect of the inductance $L$ is proportional to the magnetic flux that a current 
circulating in the loop can produce: $2\pi LI_{\mathrm{c}i}/\Phi_0$, see Eq.~(\ref{1vuo}). This effect 
can be dominating with large JJs, which can be fabricated with smaller relative parameter scatter. 
Hence the balanced SQUID does not outperform the dc SQUID in that case.

The inductances of the SQUID loops consist of geometric and kinetic inductance. The geometric
inductances were calculated with the superconducting version of \mbox{FastHenry}~\cite{fasthenry}, 
and the kinetic inductance by assuming the resistivity of the aluminum loops to be
$\rho\approx 2.7$~$\mu\Omega$cm. The total calculated inductance matrix is
\begin{equation}\label{Lmat}
\mathbf{L}/(2\pi/\Phi_0)=\left( 
  \begin{array}{cccc}
   113 & -9  & 0  & 0 \\
   115 & 124 & -5 & 0 \\
   0   & 0   & 3  & -157 \\
  \end{array}
\right) \; \textrm{pH}.
\end{equation}
Here we have neglected the cross coupling terms except $L_{21}$, which is large due to the physical 
position of current injection to the structure, as shown in Fig.~\ref{fig:sample}.

We simulated the flux modulations of the critical current with inductances from $\mathbf{L}$ of
Eq.~(3) up to $20\mathbf{L}$, and with or without the detector junction. The critical current 
suppression ratio of the balanced SQUID is about 500 with $\mathbf{L}$ and about 350 with
$2\mathbf{L}$. In the four-junction model, increasing the inductance to $5\mathbf{L}$ or higher 
breaks the simple sinusoidal modulation, and the modulation amplitude is increased. Such effects 
were not observed in our measurements, which supports the estimated values of the inductances.

Noise is expected to round off the apparent critical current suppression. An obvious noise 
source is the global magnetic field, which has been diminished by a high-$\mu$ and a superconducting 
shield. In low-$T_\mathrm{c}$ SQUIDs, the main $1/f$ noise sources are the critical-current 
fluctuations and magnetic flux noise related to defects close to the superconducting loop
lines~\cite{perf}. Our measurement is, however, relatively insensitive to magnetic flux noise, since 
the phase of the detector flux modulation turns by $2\pi$ around the minimum of the critical current 
of the balanced SQUID. Thus, magnetic flux noise creates modulation signals with all 
possible phases, and in the first order, this modulation caused by the noise averages to zero. The 
effect of flux or critical-current noise was not observed in the measurements.

In conclusion, we have demonstrated experimentally, that the critical current of the balanced SQUID
agrees well with the zero-inductance model. This is supported by theoretical estimation of the
effect of the inductances. A conservative estimate for the critical current suppression factor is 300 
or more. This number is an order of magnitude higher than the typical values obtained for
two-junction dc SQUIDs with similar maximum critical currents. Furthermore, the suppression ratio 
does not essentially depend on the exact critical currents of the individual junctions, and hence the 
fabrication yield is high.

We acknowledge the Finnish Academy of Science and Letters, V\"ais\"al\"a foundation, Technology 
Industries of Finland Centennial Foundation, and the Academy of Finland for financial support.

\end{document}